# Intrinsic Insulating State and Evidence for Phase Transitions in Ultra-clean Bilayer Graphene

Wenzhong Bao[1]*, Jairo Velasco Jr.[1]*, Fan Zhang[2,3]*, Lei Jing[1], Brian Standley[4], Dmitry Smirnov[5], Marc Bockrath[1], Allan MacDonald[2], Chun Ning Lau[1]

[1] Department of Physics and Astronomy, University of California, Riverside, CA 92521

[2] Department of Physics, University of Texas at Austin, Austin, TX 78712

[3] Department of Physics and Astronomy, University of Pennsylvania, Philadelphia, PA 19104

[4] Department of Applied Physics, California Institute of Technology, Pasadena, CA 90211

[5] National High Magnetic Field Laboratory, Tallahassee, FL 32310

*These authors contributed equally

*Abstract*

**Bilayer graphene (BLG) at the charge neutrality point (CNP) is strongly susceptible to electronic interactions, and expected to undergo a phase transition into a state with spontaneous broken symmetries. By systematically investigating a large number of singly- and doubly-gated bilayer graphene (BLG) devices, we show that an insulating state appears only in devices with high mobility and low extrinsic doping. This insulating state has an associated transition temperature $T_c$~5K and an energy gap of ~3 meV, thus strongly suggesting a gapped broken symmetry state that is destroyed by very weak disorder. The transition to the intrinsic broken symmetry state can be tuned by disorder, out-of-plane electric field, or carrier density.**

Bilayer graphene (BLG) has provided a fascinating new platform for both post-silicon electronics and exotic many-body physics [1-23]}. Because its conduction and valence bands touch at two points in momentum space and have approximately quadratic dispersion accompanied by vorticity J=2 momentum-space pseudospin textures, charge neutral BLG is likely to have a broken symmetry ground state in the absence of disorder [6-11, 15-18, 24, 25]. Theoretical work on the character of the ground state in neutral bilayer graphene has examined a variety of distinct but related pseudospin ferromagnet states, including gapped anomalous Hall states[5, 6, 19] and layer antiferromagnetic states [6-8, 16-18, 22] that break time-reversal symmetry, and gapless nematic states which alter Dirac point structure and reduce rotational symmetry [6-11, 15-18, 24, 25]. Recent experiments [19-23] have reported low-temperature conductivity $\sigma_{min}$ values for neutral BLG that vary over a large range, from ~0.05 to 250 μS. Because radically different transport characteristics are observed in samples that are apparently quite similar, it has been difficult to draw firm conclusions about the nature of the low-temperature electronic state.

Here we report on a systematic study of the minimum conductivity $\sigma_{min}$ in a large number of single-gated and double-gated BLG samples with mobilities ranging from 500 to 2000 $cm^2$/Vs for substrate-supported samples, and 6000 to 350,000 for suspended samples. We find a surprisingly constant $\sigma_{min}$ value ~ 2-3 $e^2/h$ for the majority of devices, independent of device mobility and the presence or absence of substrates (here $e$ is the electron charge and $h$ the Planck constant). However, the best devices manifest an insulating state with an energy gap ~ 2-3 meV. This bimodal distribution of $\sigma_{min}$ suggests that transport in conducting devices either arise from single particle physics, or occurs along domain boundaries that separate regions with different spontaneous quantum Hall states. Interestingly, for insulating devices, a transition between conducting and insulating states can be driven by temperature, charge density $n$, and perpendicular electric field $E_\perp$.

We fabricate single-gated BLG devices using a lithography-free technique, and suspended double-gated BLG by combining acid etching with a multi-level lithographic technique to make devices with suspended top gates [26]. All as-fabricated suspended BLG devices have relatively low mobilities, presumably due to gas or water absorption on the surface of BLG when exposed to an ambient environment. Current annealing is performed in vacuum (Fig. 1C), generally the optimal state is achieved when $I$ starts to saturate at ~ 0.2mA/µm/layer *(26)*.

Fig. 1D-E display the two terminal differential conductivity $\sigma=(L/W)dI/dV$ of two suspended BLG devices *vs.* back gate voltage $V_{bg}$ at $T$=1.5K after current annealing. Here $L/W$ is the aspect ratio of the device. Both curves are steeply *V*-shaped, with CNPs (marked by conductivity minima) close to $V_{bg}$=0V. Surprisingly, the $\sigma_{min}$ values of the devices are drastically different – 2.5 and 0.02 $e^2/h$, respectively. The insulating behavior of the latter device is confirmed by current-voltage $I$-$V_{sd}$ curves. In a magnetic field $B$, both devices display quantum Hall plateaus with the 8-fold degeneracy [12, 13] of the zero energy Landau level (LL) fully lifted*(26)*. From the Landau fan diagram that plots the differential conductance $G$ (color) *vs.* $V_{bg}$ and $B$ (Fig. 1D-E, insets), the $\nu$=0 state is visible for both devices at B>0.5T and persists down to $B$=0 for the device with very low $\sigma_{min}$ [21, 22].

To elucidate the origin of the large range of $\sigma_{min}$, we investigated 9 substrate- supported BLG devices and 23 suspended BLG devices with aspect ratios between 0.5 and 2, and areas 1-18 µm$^2$. The results are summarized in Fig. 2A, which plots $\sigma_{min}$ as a function of field effect mobility $\mu = \frac{1}{e}\frac{d\sigma}{dn}$ for each device. Evidently, the data points separate into two groups. Most data points fall into group I, in which $\sigma_{min}$ is almost independent of mobility and similar for suspended and supported devices. Within this class of devices the CNP conductivity ~ 100µS ~*2.8 $e^2/h$* [27-33].

Very different behavior is found in the 7 devices that fall into group II – $\sigma_{min}$ is at most 0.4 $e^2/h$, and as low as 1 µS. Notably, all 7 devices have very high mobility. To shed further light on the physical difference between the two groups, we also examine $V_{CNP}$, the devices' applied $V_{bg}$ at the CNP, which indicates the overall doping level. Fig. 2B-C display $\sigma_{min}$ and $\mu$ vs. $V_{CNP}$ for all suspended samples with the insulating devices denoted by blue triangles. Two striking features are evident: (1) $\mu$ decreases with increasing $V_{CNP}$ in agreement with previous reports in substrate supported graphene[34, 35], suggesting that charged impurities remain important scatterers even in these high mobility devices; (2) the insulating-BLG devices in Fig. 2B-C cluster around $V_{CNP}$=0. Thus, the insulating behavior in BLG at the CNP is only observed in devices with *both* high mobility and low charged impurity density. The insulating state, that is apparently masked by impurities in group I samples, occurs for devices with fewer scatterers and cannot be explained by single-particle physics.

To obtain further insight we compare the temperature dependences of group I and group II devices. Fig. 3A displays $\sigma_{min}$ on a logarithmic scale *vs 1/T* for 1.4≤*T*≤100 K for one non-insulating device and two different insulating BLG devices. The inset plots the same data sets $\sigma_{min}(T)$ on linear-log scales. Amazingly, for 10<*T*<100K, the $\sigma_{min}(T)$ curve of all 3 devices collapse into a single curve. This is in contrast with the previous work on single layer[36] and trilayer graphene[37, 38], which reported large sample-to-sample variation in $\sigma_{min}(T)$. Thus, the consistent behaviors among 3 devices for $T>10K$ strongly suggest that we are indeed observing intrinsic attributes of BLG.

However, the behaviors of the two types of devices start to deviate at ~ 5-7 K – $\sigma_{min}$ of the non-insulating device decreases only modestly; in contrast, the $\sigma_{min}$ of both insulating ones exhibit an abrupt change in slope and drops precipitously for $T<5K$ where the data are well-described by $\sigma_{min}(T)=A\ exp(-E_A/2k_BT)$ (here $A$ is the pre-factor, $E_A$ is the activation energy and $k_B$ is the Boltzmann constant). The best fit is obtained by using $A=17\ e^2/h$ and $E_A\sim18K$, indicating thermally activated transport over a gap $>\sim 1.6$ meV.

These data suggest the presence of a gap in insulating devices for $T<5K$. To investigate this further, we study $\sigma$ vs. source drain bias $V_{sd}$ at the CNP[21, 22](Fig. 3B). At $T=1.4K$, $\sigma$ increases precipitously when $|V_{sd}|$ increases from 0, forming a "U"-shaped profile and reaching two dramatic peaks at ±2.8 mV, and deceases again to ~ $8e^2/h$ for $|V_{sd}|>5$mV. Such a $\sigma(V_{sd})$ curve strongly resemble the density of states (DOS) for gapped phases, like superconductors, charge density waves, perhaps most pertinently the displacement field induced gapped BLG state [4, 39-41]. Since the device has symmetric coupling to both electrodes, we take the magnitude of the gap to be half of the separation between the two peaks, ~2.8 meV. This is larger than the value ~1.6 meV obtained from thermal activation measurements, but not surprising due to level broadening by, *e.g.,* impurities. Thus, the $\sigma(V_{sd})$ curves, together with the $\sigma_{min}(T)$ measurements, unequivocally establishes the presence of a low-temperature gap ~ 2-3 meV in group II BLG's spectrum.

We now examine the $\sigma(V_{sd})$ curves of the insulating device at different temperatures (Fig. 3B). When $T$ increases from 1.4 K, $\sigma_{min}$ increases, $\sigma(V_{sd})$ adopts a "V"-shaped profile, and the magnitudes of the two peaks decrease and vanish entirely at ~5K. All these observations suggest the disappearance of the gap for $T>5K$. Our data thus provide strong evidence for a finite temperature phase transition to an insulating state with a critical temperature $T_c$~5K and a gap $\Delta/k_B$ (~20-30K). The abrupt disappearance of the gap with temperature underlies many-body interaction effects, and the rough correspondence between the critical temperature and gap scales suggests that the broken symmetry is reasonably well described by mean-field theory(26).

Our data thus far suggests a $T$-dependent phase transition in charge-neutral BLG between a conducting state and an interaction-induced insulating state. The conducting state could be due to bulk two-dimensional metallic behavior, or alternatively due to transport along topologically protected edge states supported by domain walls separating regions[5] with different spin and valley dependent Chern numbers. Future experiments will be necessary to ascertain the nature of the conducting electronic state at the CNP.

An intriguing possibility is that a quantum phase transition, *i.e.* one that is tuned by parameters other than $T$ such as disorder or electric field may take place at $T=0$. To this end, we examine the $\sigma(V_{sd})$ curves of 2 conducting devices, which have mobility 140,000 and 24,000 cm$^2$/Vs, respectively, at $T=1.4K$ (Fig. 4A). Data from an insulating device is also plotted for comparison. Remarkably, $\sigma(V_{sd})$ of both conducting devices bears a striking resemblance to those of insulating BLG at higher temperatures. In particular, the device with $\mu=140,000$ cm$^2$/Vs has a "V-shaped" profile at small $V_{sd}$, elevated $\sigma_{min}$ and smaller peaks at $V_{sd}$ ~$\pm 2.5$ mV, and resembles the curve in Fig. 3B at $T$~4K. For the device with $\mu=24,000$ cm$^2$/Vs, $\sigma(V)$ is flatter without the side peaks, thus resembling the curve from the insulating device at $T$~10K. Taken together, charge disorder and temperature have similar effects on the insulating state in BLG.

Finally we examine the effect of carrier density $n$ and an applied $E_\perp$ that induces an interlayer potential difference*(26)*. In our doubly gated BLG devices we can control $n$ and $E_\perp$ independently. Several line traces of $\sigma(V_{sd})$ at $n=0$ for different values of $E_\perp$ are shown in Fig. 4bB As $E_\perp$ increases from 0 to -7mV/nm, the "U"-shaped $\sigma(V_{sd})$ curve becomes "V"-shaped, with less prominent side features and an elevated $\sigma_{min}$, i.e. the gap size appears to be diminished by $E_\perp$. For still larger fields the well-known single-particle gap of unbalanced bilayers gradually emerges [2, 4, 41]. On the other hand, the influence of total carrier density on the insulating state is extremely sharp. At $E_\perp=0$ (Fig. 4C) – a small density $n \sim 6.2 \times 10^9 \text{ cm}^{-2}$ is sufficient to significantly obscure the gapped correlated state; when $n \sim 1.2 \times 10^{10} \text{ cm}^{-2}$, the gapped feature completely vanishes and $\sigma_{min}$ reaches $\sim 5e^2/h$.

Our experimental results thus provide strong evidence for a quantum phase transition between insulating and conducting states that is tuned by $E_\perp$, $n$ or charge disorder. Indeed these transitions are expected in the mean-field theory (MFT) of gapped spontaneous quantum Hall states in BLG. These states break layer inversion symmetry in each spin-valley flavor which improves electronic correlations and induces large momentum space Berry curvatures. Consequently, there are quantized anomalous Hall conductivity contributions which change sign with the sense of flavor layer polarization of flavors. Increasing carrier density works against broken symmetry order by Pauli blocking layer polarization and by increasing screening. MFT predicts that the spontaneous quantum Hall states disappear once the carrier density is larger than $1.47 \times 10^{10} \text{ cm}^{-2}$*(26)*. in excellent agreement with our experimental findings (Fig. 4D). The role of temperature is similar to that in BCS theory of superconductivity and there is no *Anderson Theorem* to mitigate the role of disorder. In contrast to increasing $T$ and $n$, increasing $E_\perp$ reopens the gap and induces a phase transition to a fully layer-polarized state with an abrupt change of the band topology[5,18]. Such a gap is expected to increase with $E_\perp$ to be enhanced by correlations.

In summary, our systematic study of a large number of high quality BLG devices suggests that ultra-clean charge neutral BLG devices undergo a phase transition at $T_c \sim$ 5K, from a metallic state to an insulating state with an energy gap ~2-3 meV. The latter arises from electron correlation and is likely a spontaneous quantum Hall state *without* overall layer polarization[22]. Interestingly, increasing *n* or disorder has similar effects on the insulating state as temperature, suggesting that these parameters can tune continuous quantum phase transitions. Increasing $E_\perp$ with either polarity will also induce a transition into a topologically different gapped state *with* layer polarization. In the future, we expect the rich phase diagram of BLG and similar ones in ABC-stacked multilayer graphene[18,38] will provides considerable scope for future exploration.

We thank J. Jung, S. Das Sarma, O. Vafek and Jozsef Cserti for helpful discussions. This work was supported in part by UC LabFees program, NSF CAREER DMR/0748910, NSF/1106358, ONR N00014-09-1-0724, and the FENA Focus Center. C.N.L and M.B. acknowledge support by DARPA/DMEA H94003-10-2-1003. D.S. acknowledges support by NHMFL UCGP #5068. AHM and FZ were supported by Welch Foundation TBF1473 and DOE DE-FG03-02ER45958.The trenches are fabricated at UCSB Nanofabrication Facility. Part of this work was performed at NHMFL that is supported by NSF/DMR-0654118, the State of Florida, and DOE.

**Figure 1.**

**A** and **B**. False-color scanning electron micrograph of BLG device with and without top gate. Scale Bar: 2um.

**C.** $I$-$V_{sd}$ curves of a suspended BLG during current annealing.

**D-E.** Main panels and insets: $\sigma\ (V_g)$ and $G(V_{bg}, B)$ for two BLG devices with and without insulating state at CNP (T=1.5K).

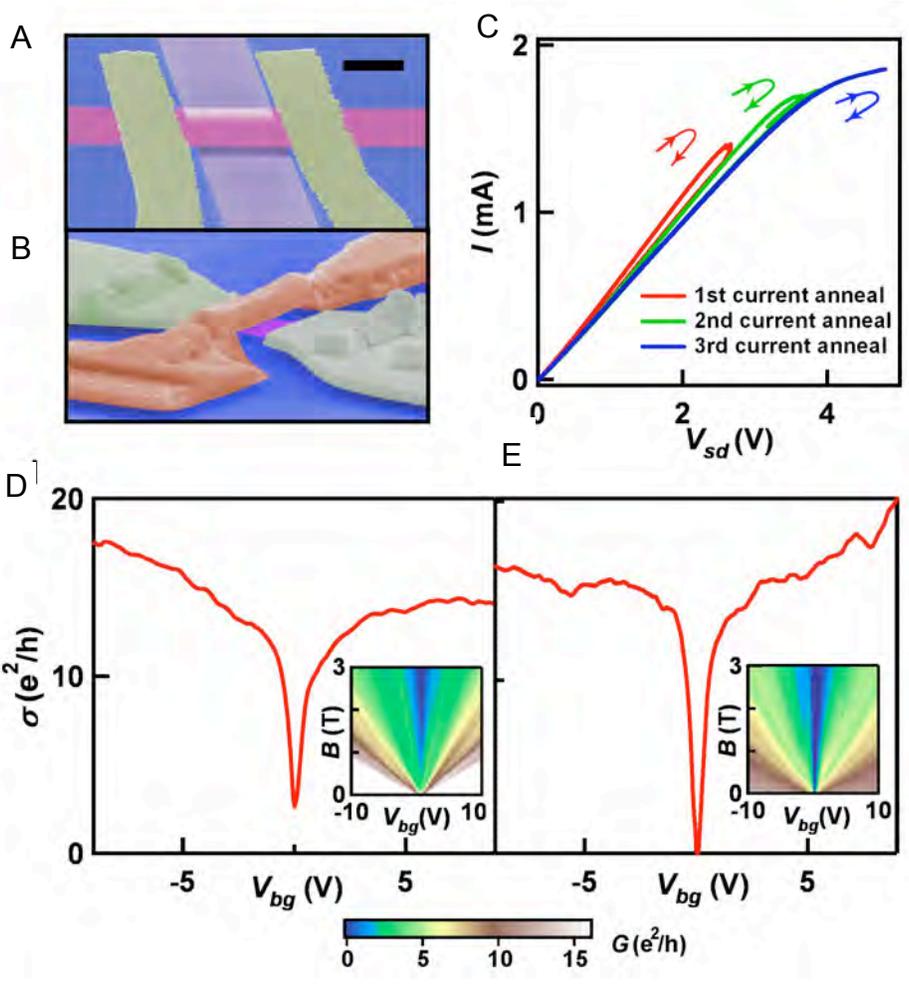

**Figure 2.**

**A.** $\sigma_{min}(\mu)$ for 9 substrate supported BLG devices (square symbols) and 23 suspended BLG devices (triangular symbols) at 1.5 K (except for one device in region 2, which was taken at *T=0.3*K*)*.

**B-C.** $\mu\ (V_{CNP})$ and $\sigma_{min}(V_{CNP})$ for suspended BLG devices. The blue symbols denote devices in region II.

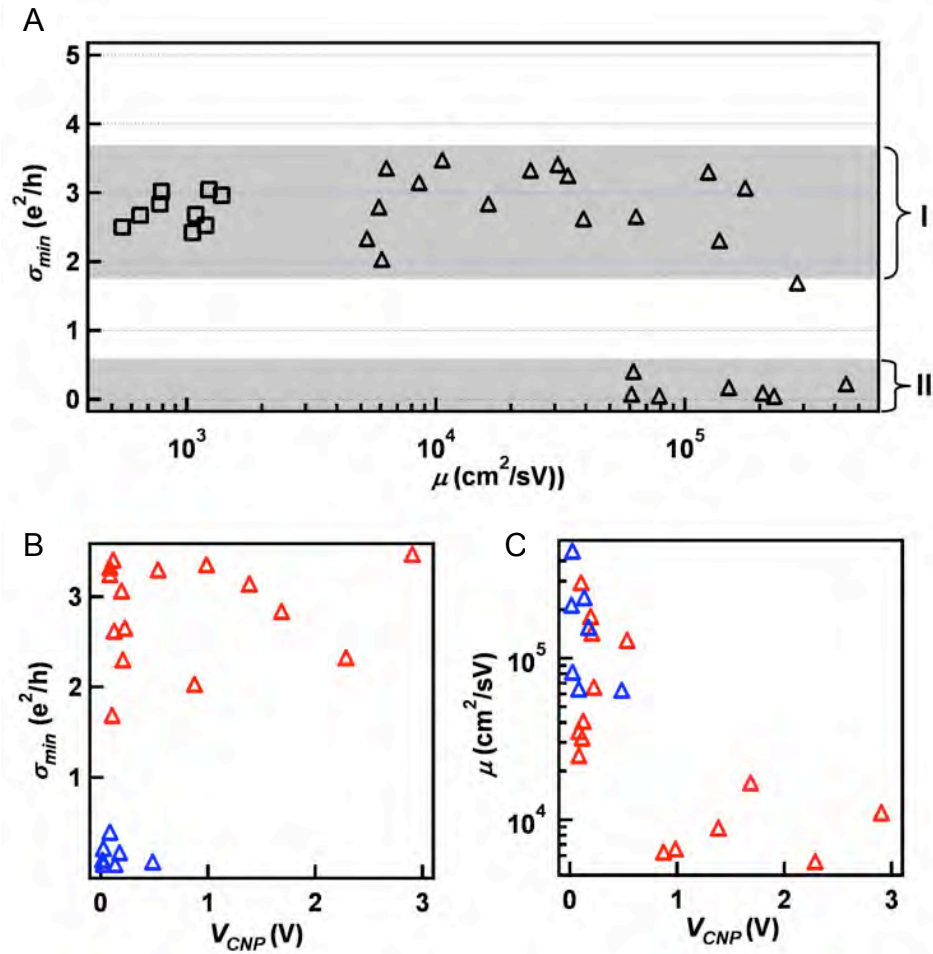

**Figure 3.**

**A.** $\sigma_{min}(1/T)$ for insulating and noninsulating-BLG devices. Inset: $\sigma_{min}(T)$ of data set. The solid lines are fits to data $T<5K$ to Eq. (1).

**B.** $T$-dependence of $\sigma(V_{sd})$ at CNP for an insulating-BLG device.

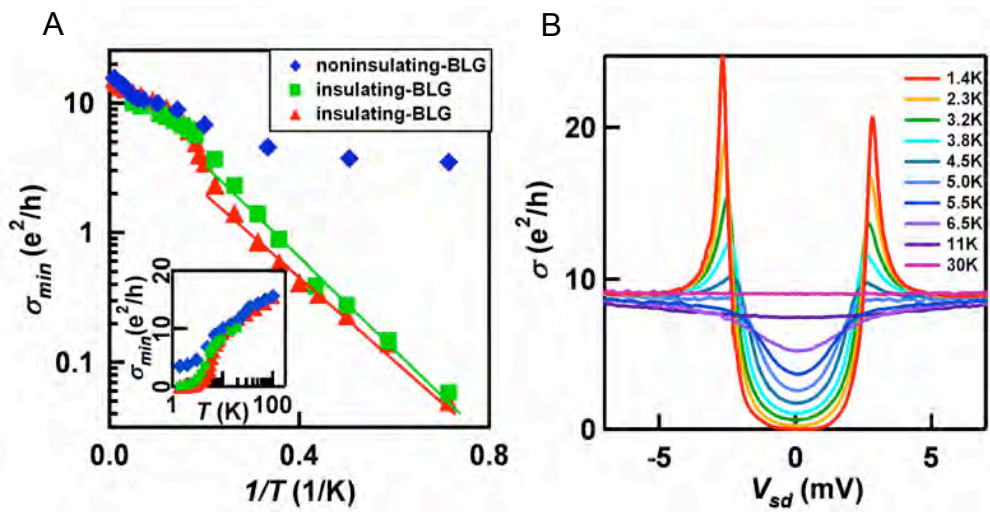

**Figure 4.**

**A.** $\sigma(V_{sd})$ for insulating and noninsulating-BLG devices at the CNP.

**B.** $\sigma(V_{sd})$ at $n=0$ for a doubly gated BLG at $E_\perp=0$, -5, -7 and -15 mV/nm.

**C.** $\sigma(V_{sd})$ at $E_\perp=0$ for a doubly gated BLG at different values of $n$.

**D.** Magnitude of flavor gap *vs.* $n$ calculated from MFT.

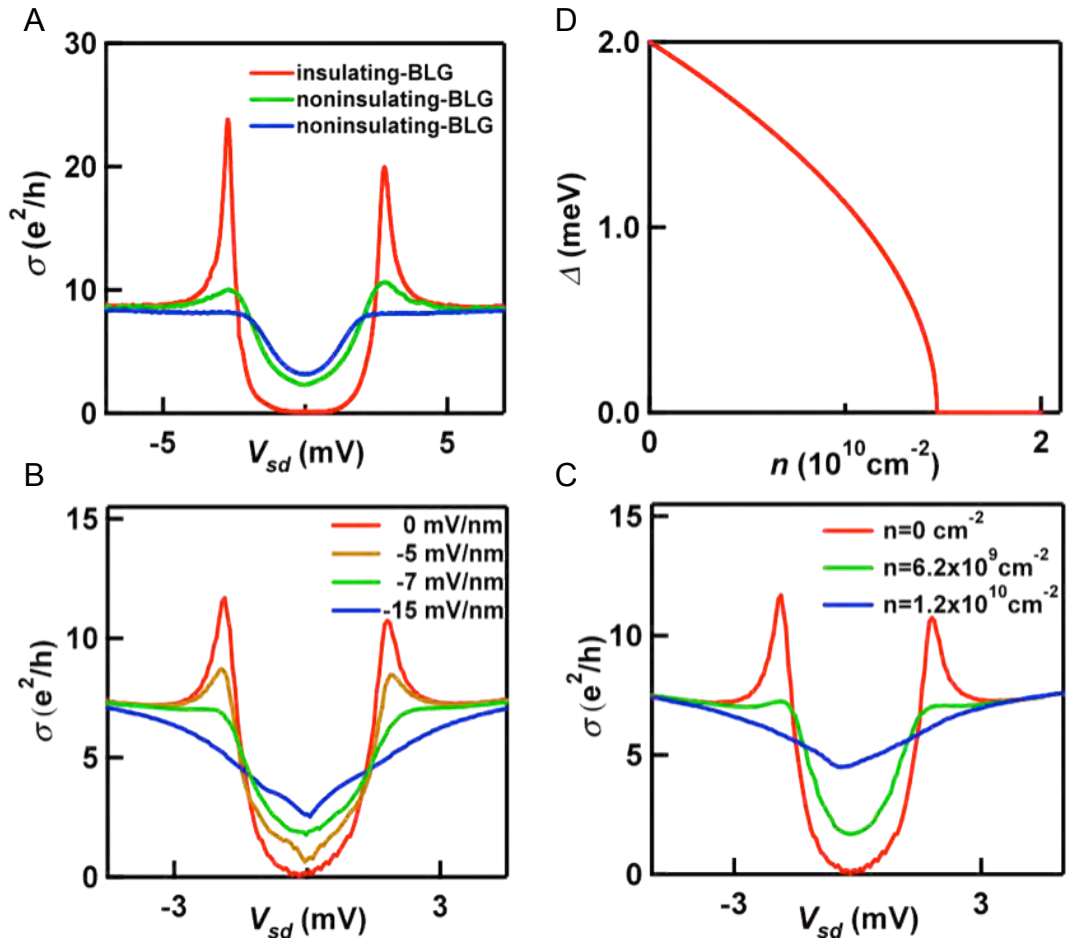